\documentclass[10pt,preprint]{aastex}
\usepackage{amssymb}
\usepackage{textcomp}
\begin{document}

\title{Most Submillimetre Galaxies are Major Mergers}

\author{H. Engel\altaffilmark{1}, L. J. Tacconi\altaffilmark{1}, R. I. Davies\altaffilmark{1}, R. Neri\altaffilmark{2}, I. Smail\altaffilmark{3}, S. C. Chapman\altaffilmark{4}, R. Genzel\altaffilmark{1}, P. Cox\altaffilmark{2}, T. R. Greve\altaffilmark{5}, R. J. Ivison\altaffilmark{6}, A. Blain\altaffilmark{7}, F. Bertoldi\altaffilmark{8}, A. Omont\altaffilmark{9}}

\altaffiltext{1}{Max Planck Institut f\"ur extraterrestrische Physik, Postfach 1312,
  85741 Garching, Germany}
\altaffiltext{2}{Institut de Radio Astronomie Millimetrique (IRAM), St. Martin d'Heres, France}

\altaffiltext{3}{Institute for Computational Cosmology, Durham University, Durham DH1 3LE, United Kingdom}
\altaffiltext{4}{Institute of Astronomy, University of Cambridge, Madingley Road, Cambridge CB3 0HA, United Kingdom}
\altaffiltext{5}{Dark Cosmology Centre, Niels Bohr Institute, University of Copenhagen, Juliane Maries Vej 30, DK-2100 Copenhagen \O, Denmark}
\altaffiltext{6}{UK Astronomy Technology Centre, Royal Observatory, Blackford Hill, Edinburgh EH9 3HJ, United Kingdom and Institute for Astronomy, University of Edinburgh, Blackford Hill, Edinburgh EH9 3HJ, United Kingdom}
\altaffiltext{7}{Astronomy 105-24, California Institute of Technology, Pasadena, CA 91125, USA}
\altaffiltext{8}{Argelander-Institut f\"ur Astronomie, Auf dem H\"ugel 71, 53121 Bonn, Germany}
\altaffiltext{9}{CNRS and Institut d'Astrophysique de Paris, 98 bis boulevard Arago, 75014 Paris, France}

\begin{abstract}
We analyse subarcsecond resolution interferometric CO line data for
twelve sub-millimetre-luminous (S$_{850\micron}$\,$\,\geq$\,5\,mJy)
galaxies with redshifts between 1 and 3, presenting new data for four
of them. Morphologically and kinematically most of the twelve systems appear 
to be major mergers. Five of them are well-resolved binary systems, and seven are compact or poorly resolved. Of the four binary systems for which mass measurements for both separate components can be
made, all have mass ratios of 1:3 or closer.
Furthermore, comparison of the ratio of compact to binary systems with that observed in local ULIRGs indicates that
at least a significant fraction of the compact SMGs must also be late-stage mergers.
In addition, the dynamical and gas masses we derive are most consistent with
the lower end of the range of stellar masses published for these
systems, favouring cosmological models in which SMGs result from mergers.
These results all point to the same conclusion, that likely most
of the bright SMGs with $L_{IR}$\,$\gtrsim$\,5$\times10^{12}L_\odot$ are
major mergers.
\end{abstract}

\section{Introduction}
\label{sec:intro}

Surveys for dust-obscured galaxies at high redshifts carried out with SCUBA \citep{holland99} and MAMBO \citep{kreysa98} revealed a high-redshift (z\,$\sim$\,1-3.5) population of sub-millimetre-luminous ($S_{850\micron}$\,$\geq$\,5\,mJy) galaxies (SMGs hereafter; see e.g.~\citealt{greve04,coppin05,chap05,pope05,weiss09}). This population consists of dusty, gas-rich, very luminous ($\sim$10$^{13}L_\odot$), intensely star-forming ($\sim$10$^3M_\odot\,yr^{-1}$) galaxies \citep[e.g.][]{greve05,chap05}.
The importance of these SMGs is underlined by studies of the
extragalactic far-IR/sub-mm background, which shows that
about half of the cosmic energy density originates in distant, dusty
starbursts and AGN \citep{puget96,pei99}, with ULIRGs contributing $\sim$20\% \citep{magnelli10}.
Near-IR \citep{swin04}, optical and X-ray \citep{alex05}, and mid-IR
\citep{lutz05,valiante07,pope08a,menen09} spectroscopic work has shown
that starbursts
are the primary power source, with AGN present in a large fraction
but not making a dominant energetic contribution.

A key question is what triggers the extraordinary luminosities of
these galaxies. A number of authors have suggested that at least some
SMGs may be interacting or merging systems, based primarily on
irregular or complex morphologies seen in rest-frame UV \textit{HST}
images \citep{chap03,chap04a,smail04,swinbank06}. However, two
\textit{caveats} need to be kept in mind regarding these results. UV
emission is strongly affected by extinction, and hence UV images may
not reliably track the underlying galaxy. This is demonstrated by the radio observations carried out by \cite{chap04a} and \cite{biggs08} (a key result of these studies was that the star formation in SMGs is more extended than in local ULIRGs, an issue which we discuss in more detail later), which show morphological differences between radio and UV emission. Furthermore, morphological
studies alone may be misleading. High-redshift galaxies are clumpy
and irregular, which may either be the result of on-going mergers
\citep{lotz06,conselice08}, or the consequence of clumpy star formation
in a kinematically more coherent structure such as an underlying
rotating disk
\citep{schreiber06,schreiber09,genzel06,elmegreen07}, or strong and structured extinction.
\cite{ivison02,ivison07} were able to show that SMGs have an increased likelihood of having close neighbours, suggestive of physical associations.
\cite{swinbank06} obtained integral field near-infrared spectroscopy data of 6 SMGs, which showed a significant fraction to consist of kinematically distinct sub-components, interpreted as an indication of merging. However, as with the \textit{HST} data discussed above and as also pointed out by \cite{swinbank10}, these rest-frame optical and UV data suffer from extinction limitations and potential biases due to outflows.
Ideally we need
spatially resolved maps and kinematics of a tracer that is not
affected by dust extinction to properly assess morphologies and
kinematics. 
Using high resolution millimetre CO line observations, 
\cite{tacc06,tacc08} find SMGs to
display either double/multiple morphologies with complex gas motions
indicative of on-going mergers, or to be extremely dense,
compact, rotating disks. The high gas and matter densities of the latter argue
for them to be end-stage major mergers. However the spatial resolution achievable with current facilities is insufficient to definitively distinguish between coalesced major
mergers and isolated disk galaxies. At present it is thus still
unclear whether SMGs form a homogeneous class of galaxies, or whether
their high luminosities can be achieved through a variety of pathways.
Illustrative of this on-going discussion are the debates about the nature of
two of the first SMGs to be discovered: SMM\,J02399-136 and SMM\,J14011+0252 \citep{ivison98,ivison00,smail02,genzel03,ivison10,downes03,smail05}.

The aim of this paper is to critically examine the hypothesis that SMGs are mergers.
We use observations of the CO line, which is unaffected by extinction or outflows,
to trace the molecular gas, and combine both morphological and
kinematic aspects in the analysis.
We present new data for four SMGs. One of our sources has not been observed in CO before (SMM\,J105141), for the three other systems these new observations provide data of significantly increased resolution. This brings the number of SMGs observed to sub-arcsecond resolution from 8 to 12 and thus provides a significant increase in the total sample of high resolution CO SMG observations. We complement these new data with previously published high resolution CO data for eight other SMGs \citep{greve05,hainline06,tacc06,tacc08,ivison10,bothwell10}, resulting in a much-improved statistical basis compared to e.g. \cite{tacc06,tacc08}. Our work is distinctive to these preceding works, the aim of which was a general characterisation of the SMG population, not only in regard to sample size, but also objective. Here we tackle a specific, heretofore unsolved key issue -- namely, whether or not SMGs are major mergers. 

In \S\ref{sec:obs} we detail our observations and data reduction. In \S\ref{sec:results} we explain the analyses employed, and present results for individual systems. We then investigate merger mass ratios, and sizes of our sample galaxies. In \S\ref{sec:disc} we test stellar mass estimates from the literature, and discuss the implication of our results for cosmological models. We then bring these results together to discuss the trigger mechanism for the exceptional luminosities of SMGs, before concluding in \S\ref{sec:sum}.

\section{Observations \& Data Reduction}
\label{sec:obs}

\begin{figure}
\begin{center}
\includegraphics[width=0.9\textwidth]{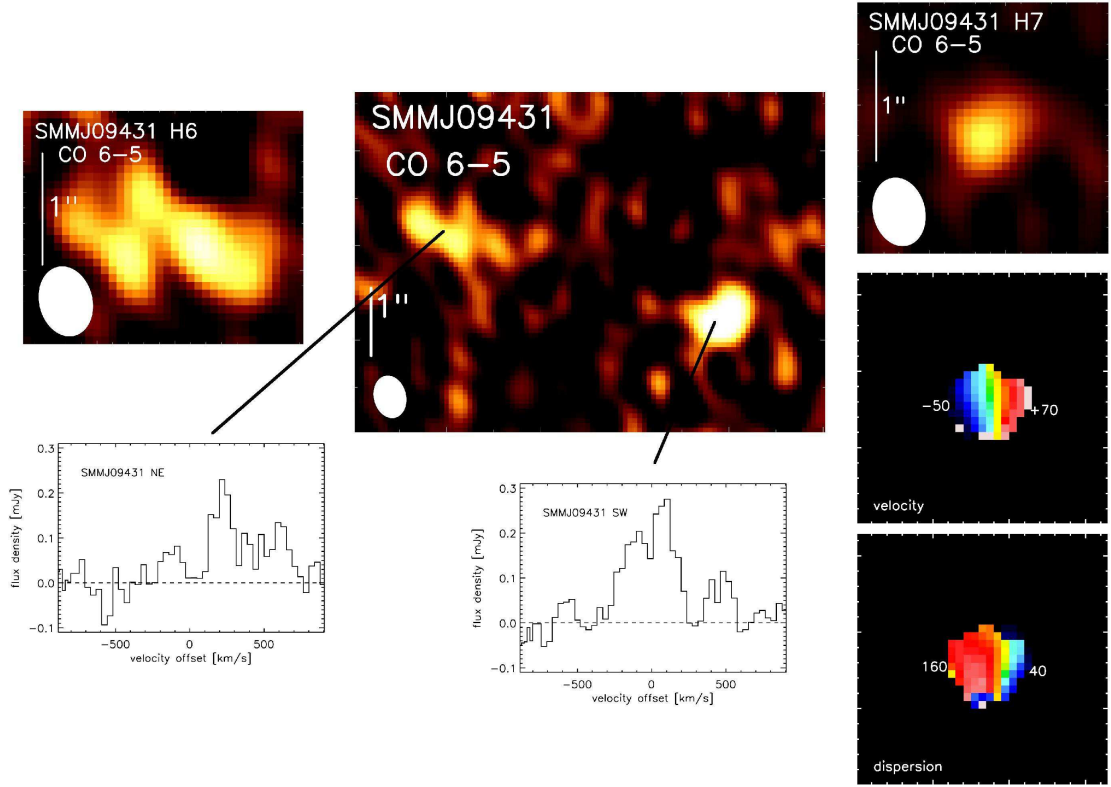}
\caption{SMM\,J09431; this system consists of two galaxies separated by $\sim$\,30\,kpc. The south-west one (H7) is a compact, rotating, dispersion-dominated disk galaxy. The left spectrum is of the north-east galaxy (H6), the right spectrum of the south-west galaxy. On the right are, top to bottom, south-west galaxy flux, velocity, and dispersion map, on the left north-east galaxy flux (the difference to the large-scale map being due to the H6 map having been integrated over the observed line extent rather than the full wavelength range) and integrated spectrum. The beam size (0.64\arcsec$\times$0.48\arcsec, P.A.\,=\,13.5\textdegree) is displayed in the lower left corner of the flux maps. North is up and east is to the left. Scale bars denote 1\arcsec.}
\label{fig:smmj09431}
\end{center}
\end{figure}

\begin{figure}
\begin{center}
\includegraphics[width=0.9\textwidth]{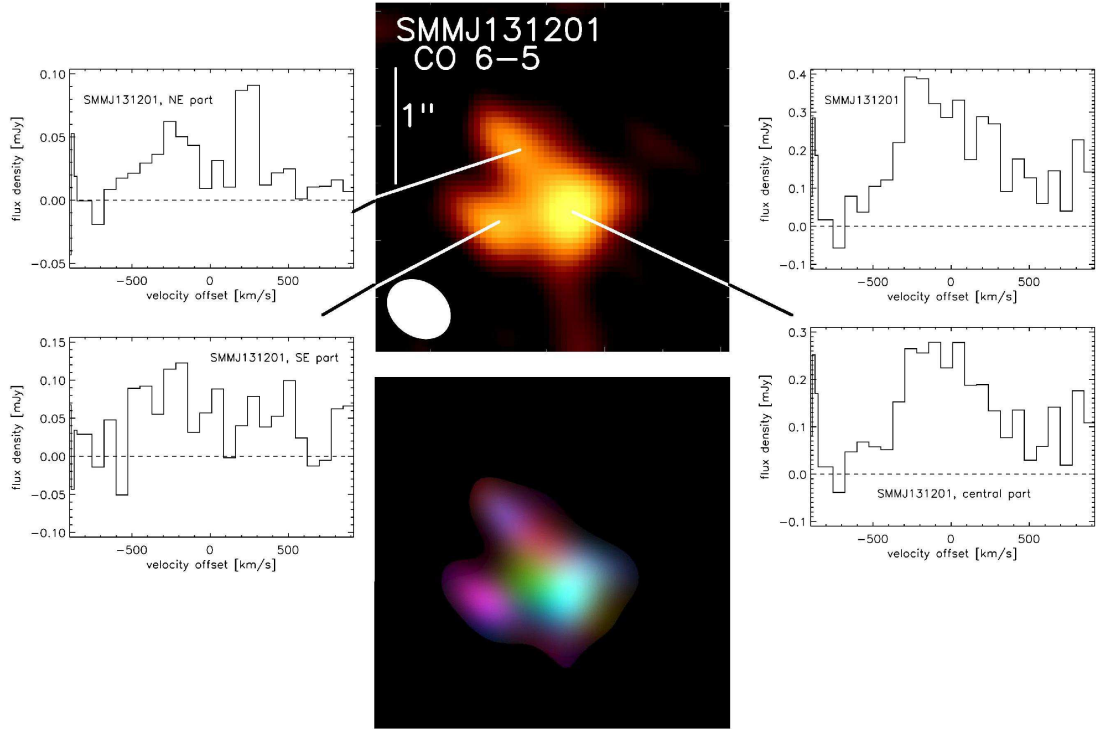}
\caption{SMM\,J131201; this system displays a complex, irregular morphology, it most likely is an advanced pre-coalescence merger. The RGB-image (mapping red, green, and blue parts of the spectrum) shows no indication of rotation. Spectra are, left: north-east (top) and south-east (bottom) arm, right: entire system (top), and central part (bottom). The beam size (0.59\arcsec$\times$0.47\arcsec, P.A.\,=\,50.9\textdegree) is displayed in the lower left corner of the flux maps. North is up and east is to the left. Scale bar denotes 1\arcsec.}
\label{fig:smmj131201}
\end{center}
\end{figure}

\begin{figure}
\begin{center}
\includegraphics[width=0.5\textwidth]{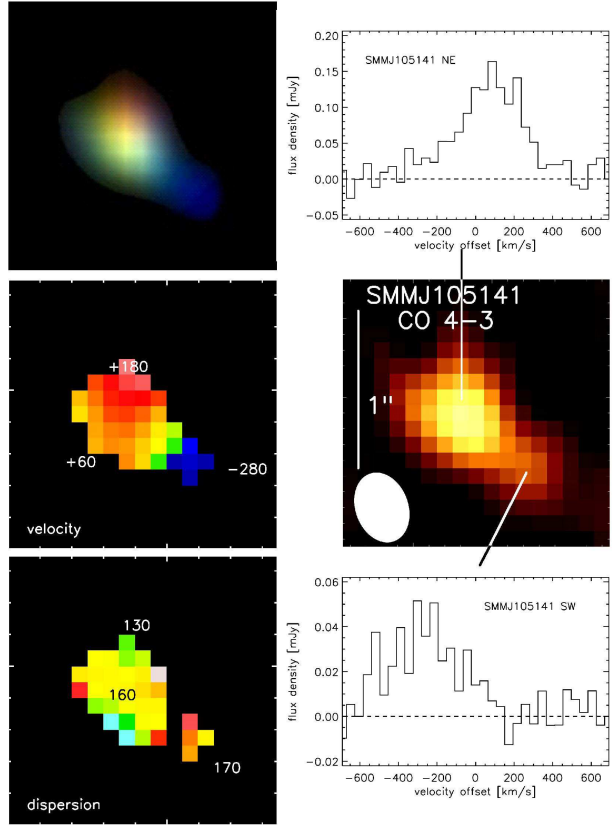}
\caption{SMM\,J105141; as the RGB-map (top; mapping red, green, and blue parts of the spectrum) clearly shows, this system consists of two galaxies that are separated in velocity by $\sim$\,400\,km\,s$^{-1}$. Top right spectrum is of `main' component (north-east galaxy), bottom right spectrum of the blue-shifted component (south-west galaxy). As can be seen from the velocity and dispersion maps (left, middle and bottom), the north-east galaxy is a dispersion-dominated disk galaxy. The beam size (0.46\arcsec$\times$0.34\arcsec, P.A.\,=\,28.7\textdegree) is displayed in the lower left corner of the flux maps. North is up and east is to the left. Scale bar denotes 1\arcsec.}
\label{fig:smmj105141}
\end{center}
\end{figure}

\begin{figure}
\begin{center}
\includegraphics[width=0.9\textwidth]{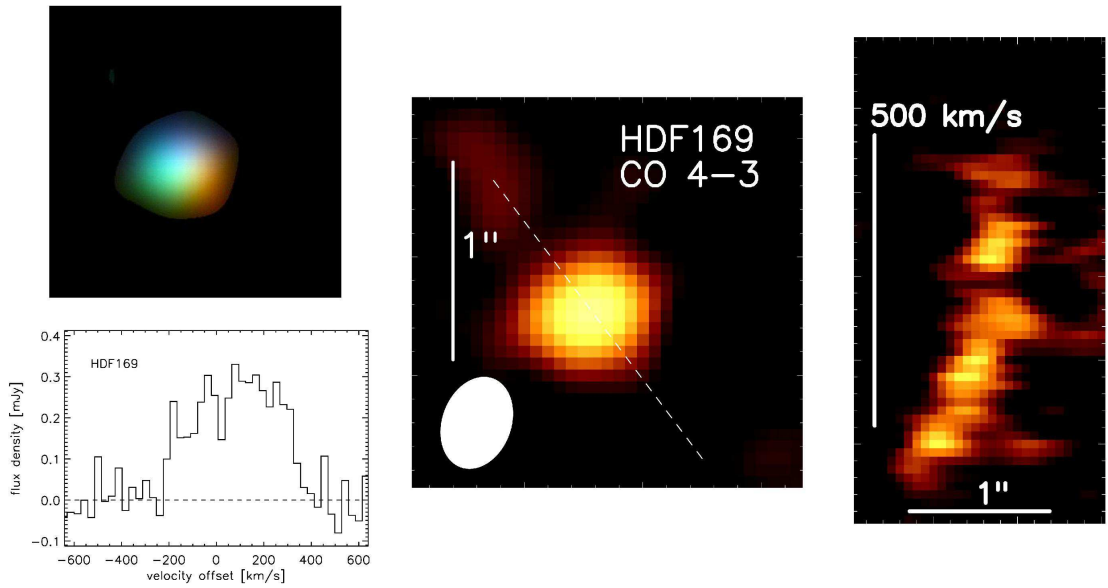}
\caption{HDF\,169; the position-velocity diagram (taken along the dotted line shown in the flux map, from lower right to upper left) and RGB-map (mapping red, green, and blue parts of the spectrum) indicate rotation, which is also supported by the integrated spectrum which appears double-peaked. This galaxy thus could either be a disk galaxy or a compact, rotating coalesced merger remnant. The beam size (0.48\arcsec$\times$0.35\arcsec, P.A.\,=\,161\textdegree) is displayed in the lower left corner of the flux maps. North is up and east is to the left. Scale bar denotes 1\arcsec.}
\label{fig:hdf169}
\end{center}
\end{figure}

The observations of four new SMGs were carried out in winter 2007/2008 with the IRAM Plateau de Bure interferometer, which consists of six 15\,m-diameter telescopes \citep{guilloteau92}. We observed the four sources, SMM\,J09431, SMM\,J105141, SMM\,J131201, and HDF\,169 under excellent weather conditions in the extended A-configuration, which has baselines up to 760 meters.  The resulting spatial resolutions ranged from 0.4\arcsec-0.6\arcsec (see Figs.~\ref{fig:smmj09431}-\ref{fig:hdf169}). For the two z\,$\sim$\,3 sources, SMM\,J09431 and SMM\,J131201, we observed the CO J\,=\,6-5 line, which is redshifted into the 2\,mm band. For the two sources at z\,$\sim$\,1.2, SMM\,J105141 and HDF\,169, we observed the CO J\,=\,4-3 line in the 1\,mm band. The observations were all made with the new generation, dual polarisation receivers, with system temperatures of $\sim$150\,K and $\sim$200\,K (SSB) in the 2\,mm and 1\,mm bands, respectively. The full available correlator capacity was configured to cover 1\,GHz in each polarisation, which adequately covers the broad CO lines seen in SMGs \citep{genzel03,neri03,greve05}. The total on-source integration times were 9.3\,hours for SMM\,J09431, 12.7\,hours for SMM\,J105141, 14.4\,hours for HDF\,169, and 20.1\,hours for SMM\,J131204.

We calibrated the datasets using the CLIC program in the IRAM GILDAS package. Passband calibration used one or more bright quasars. Every 20\,minutes we alternated source observations with a bright quasar calibrator within 15 degrees of the source. Our absolute positional accuracies are $\pm$0.2\arcsec~or better. The absolute flux scale was calibrated on MWC\,349 (1.35\,Jy at $\sim$156\,GHz and 1.6\,Jy at $\sim$208\,GHz). After flagging bad and high phase noise data, we created data cubes in the MAPPING environment of GILDAS. Final maps were cleaned with the CLARK version of CLEAN implemented in GILDAS. The absolute flux scale is better than $\pm$20\%.

\section{Results}
\label{sec:results}

Integrated spectra, integrated line flux maps, position-velocity diagrams, and velocity \& velocity dispersion maps of the individual systems are displayed in Figures~\ref{fig:smmj09431}, \ref{fig:smmj131201}, \ref{fig:smmj105141}, \& \ref{fig:hdf169}. Measured and derived source properties are listed in Tables~\ref{tab:meas}, \ref{tab:masses} \& \ref{tab:charact}.

\subsection{Analysis}
\label{sec:results:analysis}

Our analyses generally follow the procedures of
\cite{tacc06,tacc08}; for the computation of cold gas masses, we use
the approach of \cite{tacc10}.
Flux maps were produced by integrating the line emission, determining the line extent via visual inspection of spatially-integrated spectra. Fluxes are extracted from the line cubes by integrating over the area with significant emission. Given that the compact galaxies comprising a significant fraction of our sample here are in most cases only marginally resolved with the $\sim$\,0.5-1\arcsec\ spatial resolution available, we determine source sizes by fitting circular Gaussians to the systems in the \textit{uv}-plane using the task UVFIT in the IRAM data reduction package GILDAS, and taking the Gaussian FWHM$_{uv}$ to be the best estimate of the source's diameter. In the two individual galaxies of the four new systems where a Gaussian fit was not possible (most likely due to insufficient signal-to-noise), the beam FWHM was used as an upper limit to the source diameter. We note that these size measurements may slightly underestimate the actual size of the galaxies, since there may be faint extended flux below our sensitivity limit. We do not assign a formal uncertainty to this, since, if such a putative faint emission region exists, we do not know anything about its size or morphology. However the uncertainties due to this are unlikely to be significant.

Where possible, velocity and velocity dispersion maps were derived by fitting Gaussians to the line profiles.
If the spectral signal-to-noise did not permit this, we produced moment maps of different velocity ranges, which were then overlaid as RGB-images.

Since the dispersions of our galaxies are comparable to or larger than
the rotational velocities, dynamical masses have been derived using
the `isotropic virial estimator'
\citep[e.g.][]{spitzer87,tacc08,schreiber09} given in $M_\odot$ by
\begin{eqnarray}
M_{dyn,vir} = 2.8\times 10^5\Delta v^2_{FWHM} R_{1/2}
\end{eqnarray}
where $\Delta v_{FWHM}$ (km\,s$^{-1}$) is the
integrated line FWHM, and $R_{1/2}$ (kpc) is the the half-light
radius (the radius within which half the total flux is emitted, here
assumed to be equivalent to the Gaussian HWHM measured in the
\textit{uv}-plane). We note that the scaling factor appropriate for a rotating disk at an average inclination is a factor of $\sim$1.5 smaller \citep{bothwell10}, i.e.~if the above formula were applied to a disk galaxy, it would somewhat overestimate the dynamical mass. However, \cite{cappellari09} find that this estimator agrees well with masses derived from more detailed Jeans modelling for a sample of massive
z\,$\sim$\,2 early-type galaxies. 

In order to derive molecular gas masses, we convert the measured line fluxes to CO(1-0) fluxes (in Jy\,km\,s$^{-1}$) using the results of SMG SED modelling (Fig.~3 of \citealt{weiss07}, yielding $S_{CO(6-5)}$/$S_{CO(1-0)}$=13, $S_{CO(4-3)/CO(1-0)}$=10, $S_{CO(3-2)}$/$S_{CO(1-0)}$=7, $S_{CO(2-1)}$/$S_{CO(1-0)}$=3; the same numbers were also used by \citealt{bothwell10} and \citealt{casey09b}).
We assign an uncertainty of 30\% to these conversions based on the differences between the individual SED models of \cite{weiss07}; in the four
cases where we have flux measurements from two different transitions
(Table~\ref{tab:meas}) we find the flux ratios to be consistent with
these conversion factors. We do not calculate gas masses based on
CO(7-6) fluxes, since the conversion factor is uncertain for this
transition. 
We then calculate the CO(1-0) line luminosities (in K\,km\,s$^{-1}$\,pc$^2$) using Eqn.~3 of \cite{solomon97}:
\begin{eqnarray}
L'_{CO(1-0)}=3.25\times10^7 S_{CO(1-0)}\Delta V \nu_{CO(1-0),rest}^{-2}D^2_L(1+z)^{-1}
\end{eqnarray}
with $S_{CO}\Delta V$ in Jy\,km\,s$^{-1}$, $\nu_{CO(1-0),rest}$ in GHz, and $D_L$ in Mpc.
The CO(1-0) luminosities are then used to estimate the total gas mass via their Eqn.~19:
\begin{eqnarray}
M_{H_2}=\alpha L'_{CO(1-0)}
\end{eqnarray}
using $\alpha=1\,M_\odot\, / K\,km\,s^{-1}\,pc^2$ as measured for ULIRGs (\citealt{downessolomon98}, for a discussion see also \citealt{genzel10}). For an in-depth discussion of this we refer the reader to Appendix A of \cite{tacc08}; we would like to stress that our choice of $\alpha$ is motivated by the surface densities of SMGs (cf.~Table 10 of \citealt{tacc08}) rather than any \textit{a~priori} assumption about a similarity between local ULIRGs and SMGs. Our choice of conversion factor is here further supported by the fact, that if for our SMGs the Milky Way conversion factor was adopted, in all but one case the gas mass would, rather implausibly, be equal to or larger than the dynamical mass.
Finally, we apply a factor 1.36 correction to account for interstellar helium. We note that since the fluxes were measured in fixed apertures, and also some flux may be resolved out or be below our detection limit, the gas mass is likely to be somewhat larger than our estimates.

\subsection{Results For Individual Objects}
\label{sec:results:indiv}

The detailed analyses of the molecular CO line emission in
this Section reveal morphological and kinematic evidence that at least eight of the twelve SMGs studied in this paper
are on-going mergers, and the remaining four are either disk galaxies or coalesced, late-stage mergers.

\subsubsection{New Observations}

\textit{SMM\,J09431} (z\,=\,3.35) -- a review of
previous observations is given by \cite{tacc06}, who themselves presented CO(4-3) data. The system consists of two galaxies, H6 and H7,
separated by $\sim$\,30\,kpc in projection, thus physically
related and perhaps in an early stage of merging.
Our new CO(6-5) data show a clear line spectrum for the south-west galaxy H7,
with a global velocity gradient from -50 to
+70\,km\,s$^{-1}$, probably indicative of rotation. \cite{schreiber09} find that the commonly adopted boundary between `dispersion-dominated' and `rotation-dominated' systems of $v_{rot}$\,$\sim$\,$\sigma_0$ (where $v_{rot}$ is the circular/orbital velocity and $\sigma_0$ is the intrinsic velocity dispersion) translates into $\sigma_{int}$/$v_{obs}$\,$\sim$\,1.25 (where $v_{obs}$ is the observed line-of-sight rotation and $\sigma_{int}$ is the integrated line FWHM divided by 2.355). H7 has $\sigma_{int}/v_{obs}$\,$>$\,1.5 nearly everywhere, i.e.~appears dispersion-dominated. 
For the north-east galaxy H6,
we detect a 4\,$\sigma$ line shifted about +210\,km\,s$^{-1}$
relative to H7. 
Its width is $180\pm65$\,km\,s$^{-1}$. 
There may also be broader and more redshifted emission extended over a
larger region around H6 and towards H7, but at low significance. When
deriving gas masses, the measured fluxes and sizes are corrected by a factor 1.3 for
gravitational lensing effects, using the model of \cite{cowie02}.

\textit{SMM\,J131201} (z\,=\,3.41) -- This system shows complex
spatial and velocity distributions.
There is no evidence of ordered rotation, although the extensions to the
east and north-east are redshifted with respect to the emission peak.
The substructures, which may be clumps, outflows, or tidal tails, and the overall lack of rotation most likely point towards an advanced, pre-coalescence merger stage. This conclusion is also drawn by \cite{hainline08}.

\textit{SMM\,J105141} (z\,=\,1.21) -- There are two
galaxies separated in velocity by $\sim$400\,km\,s$^{-1}$ and spatially
by $\sim$5\,kpc, with no indications of disturbances. 
The south-west galaxy is about a factor 3 less
luminous than its companion, but nearly equal in dynamical mass. 
The line profiles are well-resolved. The velocity field of the
north-east system is indicative of ordered rotation, with velocities
ranging from 60 to 180\,km\,s$^{-1}$. Again we find that generally
$\sigma_{int}/v_{obs}$\,$>$\,1.5. 
It is likely that we are seeing two galaxies engaged in the early
stages of merging, appearing close only in projection.

\textit{HDF\,169} (SMM\,J123634+6212, z\,=\,1.22) -- The integrated
spectrum of this compact galaxy is double-peaked, and its P-V diagram
resembles a rotation curve, both indicative of a compact rotating
system. It has a mass surface density of
$\approx$\,$10^{3.3}$M$_\odot$\,pc$^{-2}$.
\cite{tacc08} argued that the very high volume and surface
densities characterising compact systems such as this could
only plausibly be achieved during the final stages of a major
merger. However we feel that this argument alone here is insufficient to judge with
any degree of certainty whether this system is a quiescent disk galaxy or a late-stage merger.

\subsubsection{Re-Analysis Of Existing Observations, \& Published Results}

We have re-analysed previous observations of
HDF\,242 (SMM\,J123707+614, z\,=\,2.49, \citealt{tacc06}) and N2850.4
(SMM\,J163650+4057 and SMM\,J16368+4057, z\,=\,2.38, \citealt{tacc06})
using the methods described in Section~\ref{sec:results:analysis} (see
Table~\ref{tab:meas}). 
We also complement our data with published results from CO line
observations by \cite{tacc06,tacc08} and \cite{bothwell10} for five other systems; 
HDF\,76 (SMM\,J123549+6215, z\,=\,2.20), 
N2850.2 (SMM\,J163658+4105, SMM\,J16370+4105, and SMM\,J16366+4105,
z\,=\,2.45), 
SMM\,J044307+0210 (SMM\,J04431+0210, z\,=\,2.51), HDF\,132 (SMM\,J123618.33+621550.5, z\,+\,2.00), and Lockman\,38 (SMM\,J105307+572431.4, z\,=\,1.52), where possible re-calculating derived quantities to bring them into a consistent methodological framework. 
Finally, we include results from new multi-wavelength observations of
SMM\,J02399-0136 \citep{ivison10}. 
The published results on the kinematics and morphologies of these
systems are summarised below:

\textit{SMM\,J044307+0210} -- This lensed galaxy appears compact and
has a double-peaked line profile with a FWHM of $\sim$\,350\,km\,s$^{-1}$.

\textit{HDF\,242} -- a wide binary with a separation of $\sim$\,22\,kpc.

\textit{N2850.4} -- The CO(3-2) data show a double-peaked line profile
and a separation of $\sim$\,3.3\,kpc between the peaks of the red and
blue channel maps. The integrated emission in the higher resolution
CO(7-6) data directly shows two peaks separated by
$\sim$\,3.7\,kpc. This system is therefore interpreted as a close
binary. We infer a mass ratio of 0.68 from the flux ratios of the two
components.

\textit{HDF\,76, N2850.2, \& HDF\,132} -- All compact sources with double-peaked
line profiles. They could be either coalesced major mergers, or massive disk galaxies.
They have extremely high mass surface densities of $10^{3.5}$,
$10^{4.5}$, and $10^{4.0}M_\odot$\,pc$^{-2}$ respectively \citep{tacc08,bothwell10},
at least an order of magnitude larger than those measured in the UV/optically-selected SINS galaxies \citep{schreiber06,schreiber09,tacc08} and comparable
to those measured in the central nuclear gas disk of
Arp\,220. \cite{tacc08} argue, that to achieve this, the torques
provided by a highly dissipative, major merger are necessary.

\textit{Lockman\,38} -- \cite{bothwell10} interpreted this object as a very late-stage merger, having two kinematically distinct components that are not spatially resolved.

\textit{SMM\,J02399-0136} -- New multi-wavelength observations
\citep{ivison10} reveal that this system consists of at least two major
components (most likely three, two of which host a BH);
it is therefore interpreted as an on-going merger.

\subsection{Mass Ratios}
\label{sec:results:massratios}

\cite{ivison02} first noted a preponderance of radio 
doubles among SMGs. With a larger sample, \cite{ivison07} were able to demonstrate that 
the likelihood of SMGs having companions within a few arcseconds is significantly larger than would be expected by
chance. 
This overdensity suggests that at least a significant fraction of SMGs
may currently be undergoing mergers.
We are able to refine this picture substantially by investigating the
mass ratios of the merging components.
Five of the twelve SMGs in our sample consist of spatially separated interacting galaxies, and we
are able to measure mass ratios for four of them (Table~\ref{tab:charact}; SMM\,J09431, SMM\,J105141, HDF\,242, N2850.4; no mass estimates for the individual components of SMM\,J02399 are published).
In all cases the mass ratios are consistent with being \textit{major}
(mass ratios of 1:3 or closer to unity) mergers. 
It must be kept in mind that if these are mergers between disk galaxies, the dynamical masses may be over- or underestimated due to the dependence on viewing angle, and hence the uncertainties on these mass ratios are somewhat larger than the propagated errors stated in Table~\ref{tab:charact}.
All measured spatial separations ($\lesssim$\,30\,kpc) are well within the range where they would be deemed gravitationally interacting, and the masses of individual components are $\gtrsim$10$^{10}M_\odot$, much larger than clumps in a single galaxy \citep{schreiber09}.

We furthermore note that, as \cite{bothwell10} outline, one can also make a strong case for Lockman\,38 and HDF\,132 being mergers; they show clear indication of two kinematically distinct components, which are however not individually spatially resolved. \cite{bothwell10} calculate the mass ratios of the two subcomponents to be 0.24$\pm$0.18 and 0.68$\pm$0.14, respectively. However, since these two systems cannot unambiguously be classified as mergers, we do not include them in Table~\ref{tab:charact}.



\section{Interpretation and Discussion}
\label{sec:disc}

\subsection{Sizes}
\label{subsec:sizes}

Size measurements of SMGs based on CO interferometric and mm-continuum
data have indicated that these galaxies are comparatively compact, with
median diameter $\leq$\,4\,kpc (FWHM of Gaussian
fitted to source in \textit{uv}-plane; \citealt{tacc06}). 
On the other hand, based on the diameter at the 3\,$\sigma$ detection
level of MERLIN and VLA radio observations,
\cite{chap04a} find the star formation to be `occurring on
$\sim$10\,kpc scales'.
However, \cite{tacc06} and \cite{biggs08} point out that the size measurements from the
MERLIN and VLA data are consistent with the
CO size when a comparable method (deconvolved major axes of Gaussians
fitted to the emission) is applied. 
They find a median diameter of $\sim$5\,kpc for their sample.
We can compare radio and CO size measurements individually for two sources in our sample. For SMM\,J105141 \cite{biggs08} measure an overall radio size of 0.61\arcsec$\times$0.44\arcsec, or 5.1\,kpc\,$\times$\,3.7\,kpc; we measure a CO(4-3) FWHM of 3.5\,kpc for the main component of this two-galaxy system. For HDF\,169, we find a CO(4-3) FWHM of 4.3\,kpc and measure 4.5\,kpc from an unpublished MERLIN radio map. Thus there appears to be good agreement between CO(4-3) and radio extents in these two cases.
\cite{younger08,younger10} measure the scale of far-infrared emission of four SMGs to be $\sim$5-8\,kpc. This is somewhat larger than what we find for the sample discussed here (although \citealt{younger08} say that their results are `qualitatively consistent with the results of \citealt{tacc06}'). The most plausible explanation for the difference -- beside the possibility that CO lines and far-IR continuum trace slightly different things -- lies in the fact that the \cite{younger08,younger10} SMGs probe the brightest end of the luminosity distribution, with $S_{870\micron}$\,$\gtrsim$\,15\,mJy.

For our augmented sample, which contains all SMGs for which
sub-arcsecond resolution interferometric CO observations are available (which are not strongly lensed),
we have measured FWHM sizes for 11 galaxies (measuring each galaxy
individually in the case of wide binaries).
We find average and median FWHMs of 4.5 and 4.3\,kpc, with a 1-$\sigma$ distribution of 2.8\,kpc, slightly larger than the results of \cite{tacc06}. We are unable to identify a plausible explanation for this, other than small number statistics (\citealt{tacc06} measure sizes for five systems, and upper limits for a further four; here we have size measurements for twelve systems). We also obtain upper limits of 3.4, 6.5, and 2.8\,kpc for three other sources (Table~\ref{tab:meas}). 

We note that here we are collating measurements from a range
of CO transitions, from (7-6) to (1-0), which may trace
physically different environments. 
As \cite{tacc08} elaborate, transitions up to
$J$\,$\approx$\,5-6 should all provide reasonable tracers of the
overall molecular gas. However, the extent to which, say, CO(2-1) and CO(6-5) trace the same gas, is still on-going (see also \citealt{weiss07,narayanan09a}).
That the CO(7-6) may trace denser gas might explain the comparatively
small source sizes for N2850.2 and N2850.4, and hence be at least partially responsible for the small dynamical masses we derive for these two galaxies (see also the discussion in \citealt{danielson10}). 
In the one case where we have a size measurement from two different transitions (SMM\,J09431\,SW; J\,=\,4 \& J\,=\,6), the two numbers are consistent within the uncertainties.
Restricting to size measurements from $J$\,$\leq$\,6 observations yields average and median FWHMs of 5.2 and 5.0\,kpc, with a 1-$\sigma$ distribution of 2.6\,kpc.

Thus both radio and CO measurements are consistent with SMGs
having FWHM sizes of $\sim$5\,kpc.

\subsection{Stellar Mass Measurements, and their Implications for Cosmological Models \& the Co-Evolution of Black Holes and their Hosts}
\label{subsec:stellarmass}

In this Section, we use our dynamical and gas masses to test stellar
mass estimates. This is motivated by the large systematic
uncertainties associated with stellar mass measurements, and the
implications that a reliable determination of the stellar mass 
has both for a critical assessment of competing cosmological models of SMGs, as well as for the co-evolution of central black holes and their hosts \citep[e.g.][]{hainline10}.

\subsubsection{Stellar Mass Measurements}
\label{subsub:stellarmass}

\cite{hainline08} (see also \citealt{hainline10}) derived the stellar
masses of a large sample of SMGs, employing \textit{Spitzer}
observations to cover a significantly larger spectral range than
earlier attempts \citep[e.g.][]{smail04}. The \cite{hainline08} data set was also used by \cite{micha09}
to derive stellar masses.
But despite using an identical observational data base, their results are
very different: \cite{hainline08} find a median stellar mass of
6.3-6.9\,$\times$\,10$^{10}M_\odot$, whereas \cite{micha09} arrive at
a median of 3.5\,$\times$\,10$^{11}M_\odot$. 
Therefore, testing these results is essential and the next key step
towards deriving the stellar mass content of SMGs reliably.

Table~\ref{tab:masses} brings our dynamical and gas masses
together with the stellar masses from \cite{hainline08} and
\cite{micha09} for the seven systems in common, and includes gas masses
for two more galaxies based on flux measurements of \cite{greve05}.
We also show the stellar masses derived by \cite{tacc08}.
As noted in \S\ref{sec:results:analysis}, our dynamical masses may be underestimated due to undetected faint extended flux leading to somewhat smaller $HWHM_{uv}$. However, since this effect would be small and also the gas masses correspondingly larger, this is very unlikely to qualitatively impact on our conclusions here.
We furthermore note that we have neglected both dust and dark matter in our mass budget. Dust is unlikely to contribute more than a few percent, and \cite{genzel08} find the dark matter content in the central few kpc of high-z galaxies to be 10-20\%. Thus, whilst not negligible, these will not have a significant impact on the mass budget, and we have opted not to include it here for sake of simplicity.
We have converted the stellar mass estimates from a \cite{salpeter55}
to a \cite{chabrier03} IMF by reducing them by a factor 1.7
\citep{hainline08}. 
This is supported by \cite{bastian10}, who find no conclusive
evidence that an IMF different to the locally applicable
\cite{chabrier03} or \cite{kroupa01} forms should be in place at high
redshifts.
\cite{cappellari09} also find that a \cite{chabrier03} instead of a \cite{salpeter55} IMF is required to bring the stellar masses derived for a sample of z\,$\sim$\,2 early-type galaxies via SED modelling in agreement with their dynamical masses calculated from Jeans models.

We find that $(M_{gas} + M_\star) / M_{dyn}$ for the seven galaxies for
which all three mass estimates are available, has an average (median)
of 0.8 (0.9) for the adjusted \cite{hainline08} stellar masses (with a 1-$\sigma$ distribution of 0.5);
whereas for the adjusted \cite{micha09} masses, the average (median)
is 2.1 (2.1), with a 1-$\sigma$ distribution of 1.2.
We conclude that the \cite{hainline08} stellar masses, adjusted to a
\cite{chabrier03} IMF, are consistent with our independent mass measurements (Fig.~\ref{fig:masses}).

\subsubsection{Implications for Cosmological Models}
\label{subsubsec:cosmo}

Two different theoretical approaches to reproduce the sub-millimetre galaxy
population in simulations are currently being pursued.
In semi-analytic models \citep[e.g.][]{swinbank08}, the high
sub-millimetre luminosities
are achieved through bursts of star formation induced in gas-rich
major mergers (although disk-instabilities may also
occur, \citealt{bower06}); whereas numerical
simulations have also invoked `cold accretion flows'. In the latter SMGs
are assumed to be massive galaxies sitting at the centres of large
potential wells, constantly forming stars at high rates sustained by
smooth infall and accretion of gas-rich satellites \citep[e.g.][]{dave09}.

Both models have shortcomings (such as cold accretion flows being
unable to explain the short SMG lifetimes, \citealt{tacc08}) and
it is likely that both models are fundamentally limited by our current
understanding of the role of gas physics inside the haloes. As
\cite{benson10} show, implementation of cold mode accretion into
semi-analytic models does not translate into a significant difference in
cosmic star formation history compared to previous models; this may
indicate that a more crucial aspect is how the stability of
the cold gas inside the haloes
is modelled. Equally, the assumptions underlying the numerical simulations have
not been fully tested as it is computationally
challenging to run large numbers of galaxies in these models to $z$\,=\,0 to compare
their properties to the observations of local galaxy populations  (e.g.\ luminosity
function, colour-luminosity relation)
to confirm that they reproduce these. 
Clearly, a detailed discussion of this is beyond the scope of this
paper. Instead, here we will look at the stellar masses predicted by the
current implementations of each approach, and how these compare to the
adjusted \cite{hainline08} masses.

Currently, both modelling approaches have the difficulty that if the
number density of SMGs
is matched, then with a standard IMF, they underpredict the luminosities/SFRs and
vice versa
\citep{dave09}. This
may be due to systematics in the simulations \citep{swinbank08} or the
inferred properties of observed SMGs \citep{dave09}.
A key difference between the models is the stellar mass required to
match the observed number density.
In the cold accretion flow models, SMGs are the most massive galaxies
and have a median stellar mass of $2.7\times10^{11}M_\odot$
\citep{dave09}.
In the semi-analytic models, the declining stellar mass
function, major merger rate and brief SMG phase push the predicted
stellar mass down and the median is an order of magnitude lower at
$\sim$\,$2.1^{+3.0}_{-1.0}\times10^{10}M_\odot$ \citep{swinbank08}.
In Section~\ref{subsub:stellarmass} we showed it is the lower stellar
masses, calculated by \cite{hainline08} and corrected to a
\cite{chabrier03} IMF, that are physically more plausible for SMGs.
The adjusted median stellar mass for their 64 SMGs is
3.9$\times10^{10}M_\odot$,  consistent with what is required by
semi-analytic models. We therefore conclude that,
although the associated
uncertainties are certainly non-negligible, SMG stellar mass measurements
strongly disfavour cold accretion flow models, and are compatible with
hierarchical merging models.



\subsection{What Triggers The Exceptional Luminosity of SMGs?}
\label{subsec:trigger}

It has previously been conjectured, based primarily on radio and rest-frame UV
morphologies, that the high luminosities of SMGs are triggered by
merger activity. 
Our analysis in Section~\ref{sec:results} of the CO emission shows
convincing evidence that at least a majority of our twelve SMGs are indeed mergers.
In particular, we have found that all binary SMGs where we could
derive mass ratios the merger partners are near-equal in mass,
i.e. they are \textit{major} mergers.
While it is thus beyond doubt that \textit{some} SMGs are triggered
by major mergers, the question of the nature of the compact,
non-binary SMGs comprising a significant fraction of the overall SMG
sample is yet to be settled conclusively - are they end-stage major mergers, or isolated disk galaxies?

In the following we will develop the argument that they mostly are end-stage major mergers, since a major merger will be detectable as an SMG for a significant amount of time spent in the `final coalescence' phase -- and thus that the large fraction of observed binary SMGs imply a correspondingly large number of single SMGs which are coalesced major mergers.

\begin{figure}
\begin{center}
\includegraphics[width=0.7\textwidth]{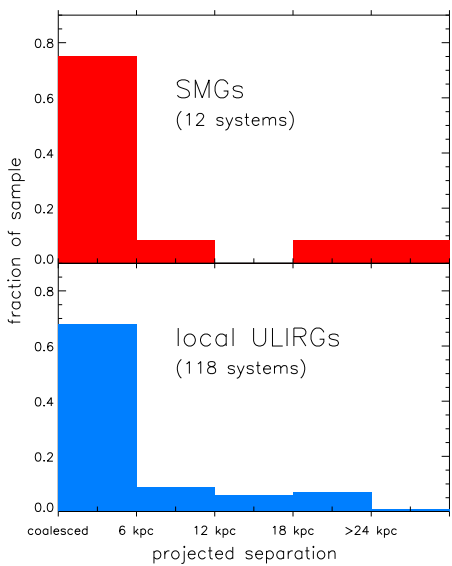}
\caption{Distribution of projected separations. Red: SMGs, this work, 12 systems. Blue: local ULIRGs, \cite{veilleux02}, 118 systems. Note that the last bin contains all systems with projected separations larger than 24\,kpc. We note that for the SMGs, of the 9 systems in the first bin (projected separation smaller than 6\,kpc), 7 are either poorly resolved or coalesced, and 2 are clearly recognisable as close binaries; the SMG sample thus divides into 5 well-resolved binary systems and 7 which are either poorly resolved or clearly compact, rotating systems. It should be noted that $\sim$\,64\% of local ULIRGs have projected separations $\leq$\,2\,kpc, however due to the larger upper limits of some of the SMGs, and the increased uncertainties of a finer resolution (due to our small sample size), we have opted to use a coarser resolution. It is evident that SMGs and local ULIRGs display a very similar distribution of projected separations.}
\label{fig:projsep}
\end{center}
\end{figure}

\begin{figure}
\begin{center}
\includegraphics[width=0.7\textwidth]{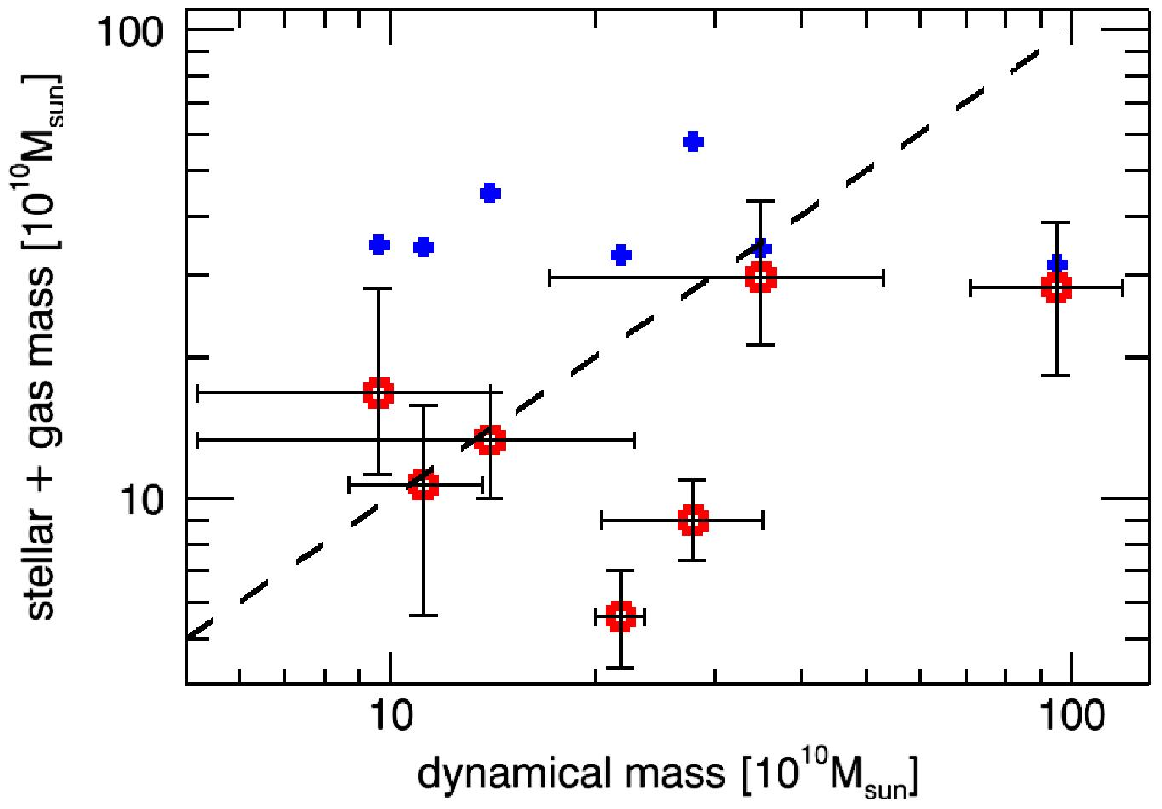}
\caption{Plot of $(M_{gas} + M_\star)$ vs $M_{dyn}$ for the seven galaxies for which all three mass estimates are available (Table~\ref{tab:masses}). Blue crosses denote \cite{micha09} stellar masses, red squares those of \cite{hainline08}; both corrected for a \cite{chabrier03} IMF. \cite{micha09} do not quote uncertainties for their stellar masses. The dashed line demarcates $M_{gas} + M_\star = M_{dyn}$; in order for the mass estimates to be physically realistic, galaxies should lie on or below this line. Within the errors, the \cite{hainline08} stellar masses pass this test, whereas those of \cite{micha09} do not. This constitutes strong support for the \cite{hainline08} results.}
\label{fig:masses}
\end{center}
\end{figure}

An illustrative comparison can be made with local ULIRGs, known to be exclusively major mergers. As Fig.~5 of \cite{veilleux02} shows, $\sim$\,64\% of local ULIRGs are either coalesced, or have projected nuclear separations $\leq$\,2\,kpc. This is what is expected from merger simulations \citep{mihos96,springel05}, which show that a first, minor starburst is triggered upon the first encounter, and the most prodigious star formation rates are only achieved during the final coalescence of the two galaxies, when tidal torquing compresses gas into the centre and leads to a $\sim$kpc-scale central starburst. Could it be that the SMG selection criterion selects a similar class of galaxies, at high redshift, and thus that the large fraction of compact SMGs in our sample is naturally explained as being those end-stage, coalesced major merger products? Indeed, as Fig.~\ref{fig:projsep} shows, local ULIRGs and our SMG sample show a strikingly similar distribution of projected separations (here we have chosen to adopt a coarser bin size than \citealt{veilleux02}, due to the comparatively small SMG sample size and the relatively large upper limits on some SMG sizes -- we note that most of the local ULIRGs in the $\leq$\,6\,kpc bin have projected separations $\leq$\,2\,kpc).
However, a direct comparison of the two samples is difficult. 
This is because $L_{IR}$ and $S_{850\mu m}$ may not always be correlated during a major merger, due to the way the infrared SED changes with dust temperature at these redshifts -- for a given $L_{IR}$, $S_{850\mu m}$ decreases with increasing dust temperature. This results in a cool dust bias in the SMG sample, illustrated by Fig.~4 of \cite{casey09a}, and borne out by observations \citep{magnelli10}. 
This is important, since it may affect which merger stages are sampled by $L_{IR}$ and $S_{850\mu m}$ selection, respectively: Both observations and simulations indicate that dust temperatures increase towards the final stages of a merger, suggesting that SMGs should sample earlier merger stages than ULIRGs. Does this imply that the observed ratio of binary and single SMGs is reconcilable with the single SMGs being isolated disk galaxies, rather than end-stage major merger products as suggested by the comparison with local ULIRGs? In order to answer this, we need to quantify the impact of the SMG cool dust bias on the merger stage sampling function.

\cite{veilleux02} use the ratio of \textit{IRAS} 25\,$\mu m$ and 60\,$\mu m$ flux as a proxy for dust temperature, and find that local ULIRGs with $f_{25}/f_{60}\geq$\,0.2 (`warm ULIRGs') are only found among system with nuclear separations below 10\,kpc (no trend is evident for smaller $f_{25}/f_{60}$, i.e.~cooler systems; their Fig.~6). However, the percentage of systems with $f_{25}/f_{60}\geq$\,0.2 is constant for all nuclear separations  $\leq$\,10\,kpc. More importantly, only 14 out of 118 systems in their sample have  $f_{25}/f_{60}\geq$\,0.2 -- i.e.~less than 10\% of ULIRGs are affected by this bias! 

Turning to simulations, Figs.~3\,\&\,4 of \cite{narayanan09a} and Fig.~1 of \cite{narayanan10} show the evolution of the star formation rate, bolometric luminosity, and $S_{850\mu m}$ during simulated gas-rich major mergers. These show that after the peak in star formation rate during final coalescence, $S_{850\mu m}$ appears to drop off more sharply than $L_{bol}$, most likely due to an increasing AGN contribution effecting an increase in the dust temperature. However, variations between different simulations make it quite difficult to quantify the time a merger has both high bolometric luminosity and low $S_{850\mu m}$ (D.~Narayanan, priv.~comm.). 

Another way to tackle this question may be to compare the AGN luminosity contributions in local ULIRGs and SMGs, since from the above we would expect SMGs to be biased against the merger phases with strong AGN contributions. \cite{veilleux09} find local ULIRG luminosities to have $\approx$\,39\% AGN contribution; estimates for SMGs are more uncertain, but span a range of 8\,-\,35\% \citep{swin04,alex05,alex08,lutz05,valiante07,pope08a,menen09}. This seems to confirm a significant bias -- however, one must bear in mind the differences between high-z and local galaxies; their absolute AGN luminosities should be comparable (similar black hole masses), but the stellar luminosities are factors of several larger in high-z galaxies (due to the larger gas fractions, \citealt{tacc10,daddi10}), and we therefore expect significantly smaller relative AGN contributions in high-z galaxies. Lower relative AGN contributions in SMGs compared to local ULIRGs are therefore not necessarily a result of a merger stage bias.

The cool dust bias of the $S_{850\mu m}$ criterion implies that there should be a population of hot dust, ultra-/hyperluminous high-z galaxies which are not detected as SMGs. And indeed, \cite{casey09a} through 70\,$\mu m$ selection find a population of z\,$>$\,1 galaxies with an average dust temperature of 52\,K, which are not detected at 850\,$\mu m$. These systems have a mean far-infrared luminosity of 1.9\,$\times$\,$10^{12}L_\odot$, only slightly lower than that of SMGs, and thus it is plausible that they consitute the hot dust extension of the SMG sample. Although uncertain, the number density of these galaxies is estimated to be approximately five times lower than that of SMGs -- implying that, if indeed all these systems are major mergers, a high-z major merger may be undetectable as an SMG for approximately 1/6th of its lifetime; or $\approx$\,25\% of the time spent in the `final coalescence' phase, if all these hot dust systems are in that stage (as one would expect, based on the arguments above). This is larger than, but consistent with, the $\sim$\,10\% of local ULIRGs showing a dust temperature\,-\,merger stage correlation.

It is clear that the merger stage sampling bias of $S_{850\mu m}$ vs.~$L_{IR}$ selection is difficult to quantify. We can only say with confidence that $S_{850\mu m}$ sampling of a major merger will likely result in a somewhat smaller fraction of coalesced systems, but it certainly will not exclude all of them.
Local major mergers, selected via $L_{IR}$, consist to $\sim$\,64\% of systems with projected nuclear separations $\leq$\,2\,kpc \citep{veilleux02}; our sample of 12 SMGs contains 7 sources which are either poorly resolved or clearly single galaxies (Table \ref{tab:charact}). If we conservatively assume that a major merger is undetectable as an SMG for 50\% (30\%) of the time spent in the `final coalescence' stage, this implies we would expect the fraction of coalesced systems in the sample to decrease from $\sim$\,64\% to about 32\% (45\%). Thus \textit{at~most} 3 (1) out of the 7 compact/coalesced sources in our sample could plausibly be isolated disk galaxies, rather than end-stage major mergers.

This corroborates the argument made by \cite{tacc06}, that the extreme surface densities of the compact SMGs can plausibly only be achieved during a major merger, and is further supported by our results of \S\ref{subsubsec:cosmo}, where we show that our confirmation of the \cite{hainline08} stellar mass estimates agrees with cosmological models in which SMGs result from major mergers.

These three independent arguments all point in one direction, namely that most, if not all, SMGs are major mergers.


\section{Summary and Conclusions}
\label{sec:sum}

Our investigation of twelve sub-arcsecond resolution interferometric CO line observations of SMGs lends strong support to a merger origin of a large majority of this class of galaxies. 
\begin{itemize}
\item
Five of our systems consist of two spatially distinct galaxies, and the remaining seven have either disturbed morphologies typical for advanced, pre-coalescence mergers or are compact, dense galaxies which plausibly are late-stage, coalesced mergers.
\item
All four systems consisting of two resolved galaxies in which the masses of the two separate components could
be derived, have mass ratios of 1:3 or closer, within the errors. This demonstrates that they are
\textit{major} mergers.

\item
Our gas and dynamical mass estimates are physically consistent with
the lower end of the range of stellar masses in the literature, those of
\cite{hainline08} adjusted to a \cite{chabrier03} IMF.
This supports hierarchical merger models, whereas it disfavours cold
accretion flow models of SMG formation.

\item
The majority ($\sim$64\%) of local infrared-luminosity-selected major mergers (ULIRGs) are coalesced or have nuclear separations $\leq$\,2\,kpc. This suggests that the large fraction of binary SMGs in our sample also implies a major merger origin for at least some of the compact/coalesced SMGs. Due to the uncertain difference in merger stage sampling of $L_{IR}\geq10^{12}L_\odot$ and $S_{850\mu m}\geq$\,5\,mJy selection, we can only estimate the corresponding ratio of binary/coalesced mergers expected in our SMG sample. However, even very conservative assumptions indicate that \textit{at most} 3 of the 7 compact/coalesced sources in our sample could plausibly be isolated disk galaxies, rather than end-stage major mergers.

\item
SMG size measurements (Gaussian FWHM) from radio and CO fluxes are both consistent with diameters $\sim$\,5\,kpc (somewhat larger than the results of \cite{tacc06} -- most plausibly due to small number statistics). These sizes are in agreement with work by \cite{menen09} and \cite{hainline09}, who find that the star formation in SMGs is more extended than the kpc-scale central starburst in local ULIRGs. 
This is consistent with what is expected from theoretical predictions and simulations of mergers with higher gas fractions \citep[e.g.][]{mihos99,narayanan09a}.
\end{itemize}

Theory and observations are only beginning to uncover how differently galaxy evolution proceeds at high redshifts, with the `cold accretion flow' framework successfully addressing many of the shortcomings of the hierarchical merging scenario. Particularly in the lower-luminosity tail of the high-z galaxy population, future observational and theoretical efforts are clearly necessary to fully clarify the picture. However, the evidence presented here overwhelmingly supports a framework in which the \textit{most} luminous z\,$\sim$\,2 galaxies are powered by major mergers.

\acknowledgements
We would like to thank the anonymous referee for a thoughtful and thorough reading of the manuscript, which helped to improve and clarify the paper. H.~Engel would like to thank Desika Narayanan for a helpful discussion of his simulations and the merger evolution of $S_{850\mu m}$ vs. $L_{IR}$.
We thank the staff of the IRAM observatory for their support of this program.
The Dark Cosmology Centre is funded by DNRF.
TRG acknowledges support from IDA. IRS acknowledges support from STFC.

\bibliographystyle{apj}

\begin{deluxetable}{lcccccc}
\tablecaption{Measured \& Derived Quantities\label{tab:meas}}
\tablewidth{0pt}
\tablehead{
\colhead{galaxy} &
\colhead{obs. CO} &
\colhead{$S_{CO}\Delta V$} &
\colhead{$HWHM_{uv}$} &
\colhead{$v_{FWHM}$} &
\colhead{$M_{gas}$} &
\colhead{$M_{dyn}$} \\
\colhead{} &
\colhead{transition} &
\colhead{[Jy\,km\,s$^{-1}$]} &
\colhead{[kpc]} &
\colhead{[km\,s$^{-1}$]} &
\colhead{[$10^{10}M_\odot$]} & 
\colhead{[$10^{10}M_\odot$]} 
}
\startdata
\bf{SMM\,J09431, SW} & \bf{6-5}\tablenotemark{1} & 1.8$\pm$0.3 & 1.13$\pm$0.23 & 390$\pm$45 & 7.10$\pm$2.44 & 4.81$\pm$1.26 \\ 
		& 4-3\tablenotemark{3} & 1.1$\pm$0.2 & 1.85$\pm$0.60 & 400$\pm$45 & 5.64$\pm$1.98 & 8.29$\pm$2.99 \\ 
		& 1-0\tablenotemark{6} & $<$0.13 & n/a & n/a & $<$8.67 & n/a \\
\bf{SMM\,J09431, NE} & \bf{6-5}\tablenotemark{1} & 1.1$\pm$0.2 & n/a & n/a & 4.34$\pm$1.53 & n/a \\
				& 4-3\tablenotemark{3} & $<$0.4 & n/a & n/a & $<$2.05 & n/a \\
\bf{SMM\,J131201} & \bf{6-5}\tablenotemark{1} & 2.54$\pm$0.1 & 4.09$\pm$0.51 & 911$\pm$140 & 13.43$\pm$4.97 & 95.04$\pm$23.81 \\ 
	 & 4-3\tablenotemark{5} & 1.7$\pm$0.3 & n/a & 530$\pm$50 & 11.67$\pm$4.06 & n/a \\ 
	 & 1-0\tablenotemark{6} & 0.42$\pm$0.07 & n/a & 1040$\pm$190 & 28.83$\pm$4.81 & n/a \\
\bf{SMM\,J105141, SW} & \bf{4-3}\tablenotemark{1} & 1.53$\pm$0.11 & $<$1.40 & 436$\pm$50 & 1.63$\pm$0.51 & $<$7.45 \\
\bf{SMM\,J105141, NE} & \bf{4-3}\tablenotemark{1} & 5.87$\pm$0.19 & 1.77$\pm$0.14 & 383$\pm$50 & 6.72$\pm$1.89 & 7.27$\pm$1.46 \\
\bf{HDF\,169} & \bf{4-3}\tablenotemark{1} & 2.25$\pm$0.11 & 2.15$\pm$0.24 & 431$\pm$60 & 2.45$\pm$0.75 & 11.18$\pm$2.53 \\
	& 2-1\tablenotemark{8} & 3.45$\pm$0.93 & $<$16.5 & 560$\pm$90 & 12.47$\pm$5.03 & $<$145 \\
HDF\,76 & 6-5\tablenotemark{2,3} & 2.3$\pm$0.4 & 0.95$\pm$0.4 & 600$\pm$80 & 5.76$\pm$2.00 & 9.58$\pm$4.42 \\
	& 3-2\tablenotemark{2,3} & 1.6$\pm$0.2 & $<$2.1 & 600$\pm$50 & 7.44$\pm$2.42 & $<$21.17 \\
HDF\,242, SW & 3-2\tablenotemark{1,2,3} & 0.59$\pm$0.08 & 2.51$\pm$0.86 & 426$\pm$60 & 3.41$\pm$1.13 & 12.75$\pm$5.10 \\
HDF\,242, NE & 3-2\tablenotemark{1,2,3} & 0.32$\pm$0.07 & 3.45$\pm$0.98 & 394$\pm$60 & 1.85$\pm$0.68 & 15.00$\pm$5.35 \\
N2850.2 & 7-6\tablenotemark{2,3} & 3.3$\pm$0.5 & 0.8$\pm$0.5 & 790$\pm$50 & n/a & 13.98$\pm$8.83 \\
	& 3-2\tablenotemark{3,5} & 1.8$\pm$0.2 & n/a & 800$\pm$50 & 10.11$\pm$3.23 & n/a \\
N2850.4, total & 3-2\tablenotemark{2,3} & 2.3$\pm$0.3 & 2.48$\pm$1.24 & 710$\pm$50 & 12.28$\pm$4.02 & 35.00$\pm$17.85 \\
N2850.4, SW & 7-6\tablenotemark{1,2,3} & 2.01$\pm$0.18 & 0.54$\pm$0.37 & 373$\pm$50 & n/a & 2.10$\pm$1.50 \\
N2850.4, NE & 7-6\tablenotemark{1,2,3} & 1.36$\pm$0.18 & $<$1.70 & 139$\pm$50 & n/a & $<$0.92 \\
SMM\,J044307 & 3-2\tablenotemark{3} & 1.4$\pm$0.2 & $<$3.26 & 350$\pm$40 & 8.20$\pm$2.73 & $<$11.18 \\
SMM\,J02399 & 1-0\tablenotemark{4} & n/a & n/a & n/a & 100$\pm$30 & n/a \\
HDF\,132 & 4-3\tablenotemark{7} & 1.27$\pm$0.14 & 2.65 & n/a & 3.49$\pm$1.12 & 21.80$\pm$1.80 \\
Lockman\,38 & 3-2\tablenotemark{7} & 0.63$\pm$0.08 & 4.75 & n/a & 1.49$\pm$0.49 & 6.20$\pm$1.10 \\

\enddata
\tablenotetext{1}{this work}
\tablenotetext{2}{\cite{tacc08}}
\tablenotetext{3}{\cite{tacc06}}
\tablenotetext{4}{\cite{ivison10}}
\tablenotetext{5}{\cite{greve05}}
\tablenotetext{6}{\cite{hainline06}}
\tablenotetext{7}{\cite{bothwell10}}
\tablenotetext{8}{\cite{frayer08}}
\tablecomments{Measured quantities for the twelve systems in our sample. New observations presented in this paper are in bold face. If a system consists of two galaxies, SW or NE indicates which galaxy is referred to. As outlined in the text, the HWHM of Gaussian fits to the source in the \textit{uv}-plane were taken as best estimates of the intrinsic $R_{1/2}$, except for HDF\,132 and Lockman\,38, for which the half light radii as quoted by \cite{bothwell10} are given. The quoted values for dynamical masses were calculated using the `isotropic virial estimator' (see \S\ref{sec:results:analysis} for details), except for HDF\,132 and Lockman\,38, for which \cite{bothwell10} estimated the quoted masses using disk rotation models. Gas masses were calculated by converting the measured line flux to CO(1-0) flux using Fig.~3 of \cite{weiss07}, assuming a ULIRG conversion factor $\alpha$\,=\,1\,$M_\odot\, / K\,km\,s^{-1}\,pc^2$, and including a factor 1.36 correction for helium.
For SMM\,J09431, measurements were corrected for gravitational lensing magnification of factor 1.3 \citep{cowie02}. 
If a source was unresolved, the beam HWHM was taken as an upper limit to $R_{1/2}$.
}
\end{deluxetable}

\begin{deluxetable}{lcccc}
\tablecaption{Merger Characteristics\label{tab:charact}}
\tablewidth{0pt}
\tablehead{
\colhead{galaxy} &
\colhead{z} &
\colhead{proj. sep.} &
\colhead{mass ratio (gas)} & 
\colhead{mass ratio (dyn)} \\
\colhead{} &
\colhead{} &
\colhead{[kpc]} &
\colhead{} & 
\colhead{}
}
\startdata
SMM\,J09431\tablenotemark{1,3} & 3.35 & 30$\pm$3 & 0.41$\pm$0.2 & n/a \\ 
SMM\,J131201\tablenotemark{1} & 3.41 & $\le$6 & n/a & n/a \\ 
SMM\,J105141\tablenotemark{1} & 1.21 & 4.9$\pm$0.5 & 0.26$\pm$0.02 & $>$0.98 \\
HDF\,169\tablenotemark{1} & 1.22 & $\le$1.5 & coal. & coal. \\
HDF\,76\tablenotemark{2,3} & 2.20 & $\le$2 & coal. & coal. \\
HDF\,242\tablenotemark{1,2,3} & 2.49 & 22$\pm$2 & 0.54$\pm$0.14 & 0.85$\pm$0.49 \\
N2850.2\tablenotemark{2,3} & 2.45 & $\le$3 & coal. & coal. \\
N2850.4\tablenotemark{1,2,3} & 2.38 & 3.7$\pm$0.4 & 0.68$\pm$0.11 & $<$0.44 \\
SMM\,J044307\tablenotemark{3} & 2.51 & $\le$4 & coal. & coal. \\
SMM\,J02399\tablenotemark{4} & 2.80 & $\sim$8 & n/a & n/a \\
HDF\,132\tablenotemark{5} & 2.00 & $\le$3 & coal. & coal. \\
Lockman\,38\tablenotemark{5} & 1.52 & $\le$4 & coal. & coal. \\
\enddata
\tablenotetext{1}{this work}
\tablenotetext{2}{\cite{tacc08}}
\tablenotetext{3}{\cite{tacc06}}
\tablenotetext{4}{\cite{ivison10}}
\tablenotetext{5}{\cite{bothwell10}}
\tablecomments{Projected separations and mass ratios. Coal. denotes coalesced system. The distribution of projected separations is also plotted in Fig.~\ref{fig:projsep}. For SMM\,J09431, the CO(6-5) data presented in this work were used to derive the gas mass ratio. For SMM\,J02399, no information on dynamical or gas masses of the subcomponents was available in the literature. As can be seen, the mass ratios are all consistent with major (mass ratio 1:3 or larger) mergers, within the error ranges.}
\end{deluxetable}

\begin{deluxetable}{lccccc}
\tablecaption{Mass Measurements\label{tab:masses}}
\tablewidth{0pt}
\tablehead{
\colhead{galaxy} &
\colhead{$M_{gas}$} &
\colhead{$M_{dyn}$} &
\colhead{$M_\star$\tablenotemark{1}} &
\colhead{$M_\star$\tablenotemark{2}} &
\colhead{$M_\star$\tablenotemark{3}} \\
\colhead{} &
\colhead{[$10^{10}M_\odot$]} & 
\colhead{[$10^{10}M_\odot$]} &
\colhead{[$10^{10}M_\odot$]} &
\colhead{[$10^{10}M_\odot$]} & 
\colhead{[$10^{10}M_\odot$]}
}
\startdata
SMM\,J131201 & 18.0$\pm$9.4 & 95.0$\pm$23.8 & n/a & 10.2$^{+4.9}_{-3.3}$ & 13.5 \\ 
HDF\,169 & 7.5$\pm$5.0 & 11.2$\pm$2.5 & n/a & 3.2$^{+1.1}_{-0.8}$ & 26.9 \\
HDF\,76 & 6.6$\pm$1.6 & 9.6$\pm$4.4 & 12$^{+4}_{-3}$ & 10.2$^{+11.1}_{-5.3}$ & 28.2 \\
HDF\,242 (SW\,+\,NE) & 5.3$\pm$1.3 & 27.8$\pm$7.4 & 12$^{+2.5}_{-4.5}$ & 3.7$^{+1.4}_{-1.0}$ & 52.4 \\
N2850.2 & 10.1$\pm$3.2 & 14.0$\pm$8.8 & 25$\pm$7.5 & 3.2$^{+1.3}_{-0.9}$ & 34.6 \\
N2850.4 (SW\,+\,NE) & 12.3$\pm$4.0 & 35.0$\pm$17.9 & 23$^{+11}_{-7}$ & 17.4$^{+12.8}_{-7.4}$ & 21.9 \\
HDF\,132 & 3.5$\pm$1.1 & 21.8$\pm$1.8 & n/a & 2.09$^{+0.9}_{-0.6}$ & 29.5 \\
SMM\,J16371+4053 & 5.3$\pm$1.1 & n/a & n/a & 4.6$^{+1.7}_{-1.3}$ & 47.8 \\
SMM\,J22174+0015 & 6.7$\pm$1.7 & n/a & n/a & 3.4$^{+3.7}_{-1.8}$ & 13.8 \\
\enddata
\tablenotetext{1}{\cite{tacc08}}
\tablenotetext{2}{\cite{hainline08}}
\tablenotetext{3}{\cite{micha09}}

\tablecomments{Stellar, gas, and dynamical mass measurements for the seven galaxies which are contained both in our sample and those of \cite{micha09} and \cite{hainline08}, and two additional galaxies contained in the latter two samples for which we calculate gas masses using fluxes measured by \cite{greve05}.  \cite{micha09} do not quote uncertainties for their stellar masses. Stellar masses were corrected for a \cite{chabrier03} IMF. Binaries are treated as one system. Where multiple transitions were observed, dynamical and gas masses are averages of those derived for the different transitions, see Table~\ref{tab:meas}. As can be seen, the stellar masses derived by \cite{micha09} are significantly larger than those of \cite{hainline08}, with only the latter being in good agreement with our gas and dynamical mass measurements.
}
\end{deluxetable}

\end{document}